\documentclass{article}
\usepackage{graphicx,epsfig}
\usepackage{graphicx,subfigure}
\usepackage{arxiv}
\usepackage{amsmath}
\usepackage{amssymb}
\usepackage[utf8]{inputenc} % allow utf-8 input
\usepackage[T1]{fontenc}    % use 8-bit T1 fonts
\usepackage{url}            % simple URL typesetting
\usepackage{booktabs}       % professional-quality tables
\usepackage{amsfonts,bm}       % blackboard math symbols
\usepackage{microtype}      % microtypography
\usepackage[mathscr]{euscript}
\usepackage[dvipsnames]{xcolor}
\pdfoutput=1
\usepackage{hyperref}
\hypersetup{
    colorlinks=true,
    linkcolor=black,
    citecolor=black,
    filecolor=black,
    urlcolor=black,
}
\makeatother
\usepackage[numbers, compress]{natbib}
\title{Instability behavior of odd viscosity-induced viscous fluid over a vibrating bed}
\author{
  Md. Mouzakkir Hossain\\
  School of Energy Science and Engineering\\
  Harbin Institute of Technology\\
  Harbin 150001, China\\
  \texttt{mouzakkir123@gmail.com} \\
\And
 Mrityunjoy Saha\\
  Department of Mathematics\\
  Indian Institute of Technology Jodhpur\\
  Rajasthan-342037, India\\
  \texttt{mrityunjoy\_saha@yahoo.com} \\
     \And
  Harekrushna Behera\\
  Center of Excellence for Ocean Engineering\\
  National Taiwan Ocean University\\
  Keelung 202301, Taiwan\\
  \texttt{hkb.math@gmail.com} \\
  \And  Sukhendu Ghosh*\\
  Department of Mathematics\\
  Indian Institute of Technology Jodhpur\\
  Rajasthan-342037, India\\
  \texttt{sukhendu.math@gmail.com} \\
}

\begin{document}
\maketitle

\begin{abstract}
The manuscript focuses on the theoretical stability analysis of the viscous liquid over a vibrating inclined rigid bed when the fluid undergoes an impact of odd viscosity. Such an impact emerges in the classical fluid owing to the broken time-reversal symmetry. The rigid bottom vibrates in streamwise and cross-stream directions. The time-dependent Orr-Sommerfeld eigenvalue problem is obtained using the normal mode approach and resolved based on the Chebyshev-collocation method and the Floquet theory. The effect of the odd viscosity coefficient on the different types of instability gravitational, subharmonic, and harmonic are identified. The gravitational instability arises in the longwave region, whereas the resonated wave instability appears in the finite wavenumber region. The gravitational instability is generated in the fluid flow owing to the gravity driving force, whereas the subharmonic instability appears for lower forcing amplitude and the harmonic instability emerges for comparatively higher forcing amplitude.  
It is found that the subharmonic and harmonic resonances appear when the forcing amplitude surpasses its critical value. A higher odd viscosity leads to stabilizing the gravitational instability, whereas a larger odd viscosity diminishes the subharmonic resonance along with the harmonic resonance instigated at a high forcing amplitude. Further, a new instability, named shear instability in the finite wavenumber range emerges together with the aforementioned three instabilities when the Reynolds number is sufficiently high with a low angle of inclination and becomes weaker when the time-reversal symmetry breaks.

\end{abstract}

% keywords can be removed

\section{Introduction}

 The hydrodynamic instability of fluid flow over a vibrating plane is of great importance in various applications within atomization technology, such as the formation of fuel sprays, high-tech surface cleaning, and advanced material processing \cite{woods1995instability}. In most circumstances, the bottom vibration is built purposefully, while in other applications, the bottom vibration simply cannot be ignored. Thus, in either situation, it is of interest to us to investigate how the alterations from the ideal condition of a static plane affect the liquid film flow with various physical effects. The bottom vibration can parametrically induce the generation of so-called Faraday waves. It also has the potential to induce Stokes-Rayleigh waves in the viscous liquid layer. It was \citet{yih1968instability}, who initiated an investigation on Newtonian liquid flow over a bounding wall that oscillates horizontally. He used the standard perturbation technique within the framework of Floquet theory to solve the time-dependent Orr–Sommerfeld (OS) boundary value problem (BVP) and examined the behavior of longwave modes under the influence of unsteady base flow. He observed instability of the longwave mode when the amplitude of oscillation reaches a sufficiently large level. Following the work of \citet{yih1968instability}, an extensive examination of the literature (\cite{dandapat1975instability}, \cite{lin1996suppression}, \cite{or1997finite}, \cite{or1998thermocapillary}, \cite{burya2001stability}, \cite{gao2006effect}, \cite{samanta2009effect}, and the citations therein) explored film flow over vibrating inclined/vertical/horizontal substrates, encompassing diverse effects. \citet{or1997finite} extended the idea of \citet{yih1968instability} by investigating the finite wavelength instability for infinitesimal perturbations with an arbitrary wavenumber. His findings indicated that the unstable zone in the longwave region emerges in the isolated bandwidth of the applied frequency, whereas the instability in the finite wavenumber arises through the branch points of the longwave marginal curves. Also, he identified relatively stronger instability in finite wavenumber mode than in longwave mode in certain ranges of the imposed frequency. Later, \citet{gao2006effect} used a strategy to control instability by adding an insoluble surfactant to the surface of the liquid in the same flow model considered by \citet{yih1968instability}. They observed the stabilizing impact of insoluble surfactant on the longwave mode. Instead of considering only the streamwise-directed oscillation of a plane,
\citet{woods1995instability, lin1996suppression}, and \citet{burya2001stability} examined the instability of a Newtonian fluid flowing over an inclined plane with oscillatory motion perpendicular to its own plane, and the resulting acceleration has two components
in the streamwise and cross-streamwise flow directions. Consequently, the effective modulation of gravity causes Faraday waves on the liquid surface, where both subharmonic and synchronous solutions occur.

\citet{dandapat1975instability} extended the unsteady Newtonian flow model of \citet{yih1968instability} by choosing an unsteady non-Newtonian viscoelastic flow model. The purpose of their extension was to investigate the linear stability analysis in the longwave zone. Later, \citet{samanta2017linear} revisited the work of \citet{dandapat1975instability} and performed a detailed discussion on the linear stability for the disturbances of an arbitrary wavenumber. He observed that the finite wavenumber instability is higher in comparison to the longwave instability for some ranges of imposed frequency.
% He observed that the finite wavenumber instability is more dangerous than the longwave instability in some ranges of imposed frequency.
Recently, the work proposed by \citet{woods1995instability} was further extended by \citet{samanta2021instability} for arbitrary wave numbers by applying shear force at the surface of the liquid layer based on the idea of \citet{smith1982instability} and \citet{smith1990mechanism}. The inclined substrate beneath the shear-imposed liquid layer was considered to oscillate in streamwise and cross-stream directions. His numerical findings revealed the presence of three distinct instabilities, namely, gravitational, subharmonic, and harmonic instabilities. For small Reynolds numbers, the subharmonic resonance instigated by a low forcing amplitude becomes stronger, but it can be weakened using a high forcing amplitude while imposed shear is present. However, the externally applied shear can enhance the harmonic resonance that is triggered solely at high forcing amplitude. His motivation behind an imposing external shear at the free surface arose owing to its huge application in many industrial and natural set-ups. For instance, inside an aero-engine bearing chamber, the shearing airflow generated by high-speed rotating parts inside the chamber plays a crucial role in extracting heat from the surface of the oil film \cite{sivapuratharasu2016inertial}.

It should be noted that all the above-mentioned articles considered the even viscosity of the viscous stress tensor and avoided the odd viscosity term. However, an odd viscosity term plays a crucial role in the dynamics and instability mechanisms of the flow system. \citet{avron1998odd} identified that when the time-reversal symmetry of a viscous fluid is spontaneously disrupted or influenced by external fields, the fluid exhibits a second viscosity coefficient referred to as odd viscosity or Hall viscosity. This coefficient, in addition to the conventional viscosity coefficient, introduces an off-diagonal contribution to the Cauchy stress tensor.  Hence, it is crucial to use modified Navier-Stokes equations to accurately depict how the presence of the odd viscosity coefficient affects fluid flow dynamics. This coefficient is anticipated to have a significant impact on the complex wave dynamics exhibited by such fluids. Because of this, \citet{lapa2014swimming} pioneered a theoretical analysis of swimming strategy in two-dimensional fluids with an odd viscosity coefficient at a low Reynolds number. Later, the impact of the odd viscosity coefficient on the thermocapillary instability of a viscous film over a uniformly heated plane was explored by  \citet{kirkinis2019odd}.  They noticed the mitigation of thermocapillary instability when the odd viscosity is introduced in the liquid film.
% They noticed that the odd viscosity possesses the potential to mitigate the thermocapillary instability of the liquid film. 
\citet{bao2021odd} investigated the influence of odd viscosity on the instability of a falling liquid film in the presence of an external electric field at the liquid surface. Their analytical results acquired by the longwave expansion method revealed that the odd viscosity stabilizes the flow, while the electric
field destabilizes the flow. Recently, a comprehensive study into the linear and nonlinear stability of falling liquid film was conducted by \citet{samanta2022role} when the time-reversal symmetry was disrupted. He solved the corresponding OS BVP using the numerical technique Chebyshev spectral collocation and affirmed the stabilizing behavior of odd viscosity on the surface and shear waves of the flow. Further, a coupled system of two-equation models in terms of film thickness and flow rate was derived, and the corresponding nonlinear traveling wave solution affirmed that the presence of odd viscosity delays the transition from the base flow to the secondary flow arising from the interaction between nonlinear waves. In recent times, the exploration of the influence of odd viscosity on diverse fluid flow setups has been a focal point for several investigators \cite{chattopadhyay2021odd, paul2023hydrodynamic, hossain2024odd}. Their collective efforts advance our understanding of the complex role that odd viscosity plays in various fluid-dynamic scenarios.

Thus, in the current manuscript, our main objective is to extend the earlier work of \citet{woods1995instability} by detecting the influence of odd viscosity on the instability mechanisms of Faraday waves emerging on the surface of viscous fluid over a vibrating inclined plane for arbitrary wavenumbers. So far, no analyses have enlightened the finite wavenumber linear stability of viscous fluid with disrupted time-reversal symmetry over a vibrating substrate.
% However, so far, no attempt has been made to decipher the finite wavenumber linear stability analysis corresponding to a viscous fluid with broken time-reversal symmetry over a vibrating substrate. 
   % it is crucial 
% In this article, the main purpose is to fill the gap that remained in the previous study
% (\citet{samanta2021instability}), i.e., to identify the effect of odd viscosity on the dynamics and instability mechanisms of Faraday waves that emerged on the surface of a viscous fluid flows over a vibrating inclined plane for arbitrary wavenumbers. 
Here, the vibration of the bottom substrate is assumed to occur in both streamwise and cross-streamwise directions, with identical frequencies. Numerous methodologies have recently emerged for the implementation of laboratory-on-a-chip applications using microfluidic devices designed for fluid transport. One potential technique involves fluid movement through the application of oscillations of the bounding wall with the breaking of time-reversal symmetry. Thus, the current flow model can be regarded as a prospective approach for the development of microfluidic devices. Besides, utilizing the patterns generated by Faraday waves allows for the creation of structured surfaces, offering applications in the fabrication of patterned materials. These materials find practical use in fields such as optics, sensors, and other technological applications.

The modified time-dependent OS BVP of the current fluid flow model is obtained by applying the classical normal mode approach to the linear perturbation equations imposed on the unsteady base flow solution. The time-dependent OS BVP is numerically solved utilizing the Chebyshev spectral collocation within the framework of Floquet theory.
The following is the manuscript's layout: Section~\ref{MF} delineates the governing equations associated with the boundary conditions. Next,
Section~\ref{BVP} contains the derivation of the time-dependent OS BVP including the numerical approach. A detailed discussion of the numerical outcomes is introduced in Section~\ref{NR}, and finally, a conclusion is made in Section~\ref{CON}.

\section{Mathematical Formulation}  \label{MF}

To formulate the equations of the problem in the Cartesian coordinate system, the origin $O$ is considered at the midway of the undisturbed liquid layer of thickness $2d$. The coordinates $x$ and  $y$ are selected along the flow direction and the vertically upwards direction, respectively.
% To formulate the equations of the problem, we choose a reference frame in which the origin $O$ is taken at mid-depth of the unperturbed liquid layer of thickness $2d$, whereas the $x$ and $y$-axes are taken along and perpendicular to the inclined plane, respectively.
The bounding wall is assumed to have sinusoidal oscillation in both directions streamwise and cross-streamwise with unique frequency $\omega$.  % Suppose that the bounding plane is forced to oscillate sinusoidally with identical frequency $\omega$ in the streamwise and cross-streamwise directions.
Moreover, the viscous fluid disrupts the time-reversal symmetry, which introduces an additional viscosity coefficient, known as odd or Hall viscosity, to the Cauchy stress tensor.
Henceforth, the Cauchy stress tensor $ \mathcal{T}$ is of the following form:
\begin{equation}
    \mathcal{T}= \mathcal{T}^{e}+\mathcal{T}^{o}, \label{eq1}
\end{equation}
where the usual (even) part $ \mathcal{T}^{e}$ corresponds to the typical (even) viscosity $\mu_{e}$, while the odd part $ \mathcal{T}^{o}$
is associated with the odd (Hall) viscosity $\mu_o$ appeared owing to the breaking of time-reversal symmetry. 
% where $ \mathcal{T}^{e}$ defines the usual (even) component of the Cauchy stress tensor with the typical (even) viscosity $\mu_{e}$ and $ \mathcal{T}^{o}$ represents the odd component of the Cauchy stress tensor with the odd (Hall) viscosity $\mu_o$ that arises when time-reversal symmetry-breaking emerges. 

The above Eq.~\eqref{eq1} can be represented \cite{avron1998odd, samanta2022role} as:
\begin{align}
& \mathcal{T}^{e}_{ij}=-p\delta_{ij}+\mu_{e}\left( \partial_{x_j}u_i +\partial_{x_i}u_j \right) ~~~i,j=1,2, \label{even}
\\
& \mathcal{T}^{o}_{ij}=-\mu_{o}\left[ (\delta_{i1}\delta_{j1}-\delta_{i2}\delta_{j2} )\left( \partial_{x_2}u_1 + \partial_{x_1}u_2\right)-(\delta_{i1}\delta_{j2}+\delta_{i2}\delta_{j1} )\left( \partial_{x_1}u_1 - \partial_{x_2}u_2 \right)\right] ~~~i,j=1,2,\label{odd}
\end{align}
where $\delta_{ij}$ indicates the Kronecker delta with $x_i-$directed velocity component $u_i$. The fluid flow model is specified by choosing $(x_1, x_2)$  and $(u_1,u_2)$ as $(x,y)$ and $(u,v)$, respectively. 
% where $\delta_{ij}$ refers to the Kronecker delta. Here $u_i$ is the velocity component in the $x_i$ direction.
% In this problem, $(x_1, x_2)$  and $(u_1,u_2)$ are chosen as $(x,y)$ and $(u,v)$, respectively.
\begin{figure}[ht!]
    \centering
   \includegraphics[width=12cm]{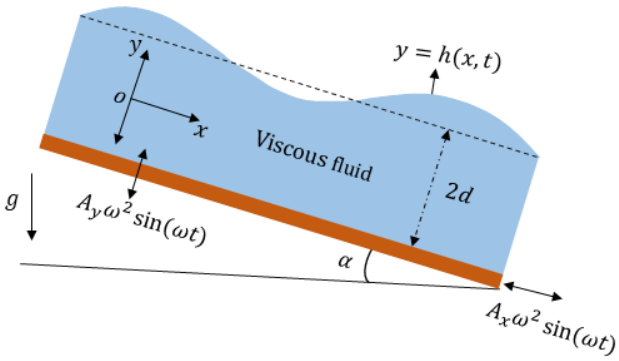} 
    \caption{Geometry of gravity-driven liquid over a vibrating plane with broken time-reversal symmetry.}
    \label{fig1}.
\end{figure}

In the Cartesian coordinate systems, the continuity and modified Navier-Stokes equations representing the motion of an odd viscosity-induced viscous and incompressible fluid over an inclined vibrating wall can be written in the form:
\begin{align}
&\partial_{x}u + \partial_{y}v=0, \label{e1}\\
&\rho(\partial_{t}u+u~\partial_{x}u+v~\partial_{y}u)=-\partial_{x}p+\mu^{e}(\partial_{xx}u+\partial_{yy}u)-\mu^{o}(\partial_{xx}v+\partial_{yy}v)+\rho g\sin{\alpha}-\rho\,A_x\,\omega^2\sin{(\omega\,t)}, \label{e2}\\
&\rho(\partial_{t}v+u~\partial_{x}v+v~\partial_{y}v)=-\partial_{y}p+\mu^{e}(\partial_{xx}v+\partial_{yy}v)-\mu^{o}(\partial_{xx}u+\partial_{yy}u)-\rho\,g\cos{\alpha}-\rho\,A_y\,\omega^2\sin{(\omega\,t)}, \label{e3}
\end{align}
where $g$ signifies the acceleration due to gravity and $p$ represents the fluid pressure. Due to the vibration of the bottom wall in the streamwise and cross-streamwise directions, extra acceleration terms $A_x\,\omega^2\sin{(\omega\,t)}$ and $A_y\,\omega^2\sin{(\omega\,t)}$ related to the d'Alembert's body force are included in the  $x-$ and $y-$ momentum Eqs. \ref{e2} and \ref{e3}, respectively,  where $A_x$ and $A_y$ denote the forcing amplitudes of streamwise and cross-streamwise oscillations, respectively.

The kinematic boundary condition at the fluid surface yields
\begin{align}
    &v=\partial_{t}h+u~\partial_{x}h\quad \text{at}\quad y=h(x,t),\label{ee4}
\end{align}
The dynamic condition possesses a stress balance at the fluid surface. The form of tangential stress at the free surface \cite{wei2005effect, samanta2014shear, samanta2022role} is
\begin{align}
&\hspace{-1cm}\mu^{e}\biggl[\biggl\{1-(\partial_{x}h)^2\biggr\}(\partial_{y}u+\partial_{x}v)-2(\partial_{x}u-\partial_{y}v)\partial_{x}h\biggr]+\mu^{o}\biggl[(\partial_{x}u-\partial_{y}v)\biggl\{1-(\partial_{x}h)^2\biggr\}+2(\partial_{y}u+\partial_{x}v)\partial_{x}h\biggr]=0\nonumber\\
&\hspace{12cm}\text{at}\quad y=h(x,t).
\end{align}
The normal stress jump boundary condition \cite{anjalaiah2013thin, bhat2020linear, sani2020effect} at the liquid surface yields 
 \begin{align}
&p-p_a=\frac{1}{1+(\partial_{x}h)^2}\biggl[2\mu^{e}\left\{\partial_{x}u\,(\partial_{x}h)^2-(\partial_{y}u+\partial_{x}v)\partial_{x}h+\partial_{y}v\right\}+\mu^{o}\biggl\{(\partial_{y}u+\partial_{x}v)\biggl\{1-(\partial_{x}h)^2\biggr\}
\nonumber\\
&\hspace{4cm}-2(\partial_{x}u-\partial_{y}v)\partial_{x}h\biggr\}\biggr] -\sigma\frac{\partial_{xx}h}{\biggl\{1+(\partial_{x}h)^2\biggr\}^{\frac{3}{2}}}\quad\text{at}\quad  y=h(x,t),
\end{align}
where $p_a$ denotes the constant atmospheric pressure and $\sigma$ denotes the surface tension.
The boundary conditions are the no-slip and no-penetration on the rigid plane, i.e.,
\begin{align}
    &u=0\quad\text{and} \quad v=0\quad\text{at}\quad  y=-d.\label{e10}
\end{align}
The dimensionless forms of Eqs.~\ref{e1}-\ref{e10} are obtained using the following dimensionless variables:
\begin{align}
& \Bar{x}=\frac{x}{d}, \quad \Bar{y}=\frac{y}{d},\quad \Bar{u}=\frac{u}{\omega d},\quad \Bar{v}=\frac{v}{\omega d},\quad \Bar{t}=t\omega,\quad\Bar{p}=\frac{p}{\rho\omega^2d^2},\quad\Bar{h}=\frac{h}{d},\quad a_x=\frac{A_x}{d},\nonumber\\
&a_y=\frac{A_y}{d},\quad \mu_H=\frac{\mu^{o}}{\mu^{e}}\label{e11}.
\end{align}

Upon substituting the Eq.~\eqref{e11} into Eqs.~\eqref{e1}-\eqref{e10}, the resulting governing equations, and the boundary conditions (after suppressing the over
bars) in the dimensionless form can be expressed as
\allowdisplaybreaks
\begin{align}
&\partial_{x}u + \partial_{y}v=0,\label{2.8}\\
&Re\bigg[\partial_{t}u+u~\partial_{x}u+v~\partial_{y}u\bigg]=-Re\,\partial_{x}p+(\partial_{xx}u+\partial_{yy}u)-\mu_H(\partial_{xx}v+\partial_{yy}v)-Re\, a_x\sin{t},
\\
&Re\bigg[\partial_{t}v+u~\partial_{x}v+v~\partial_{y}v\bigg]=-Re\,\partial_{y}p+(\partial_{xx}v+\partial_{yy}v)+\mu_H(\partial_{xx}u+\partial_{yy}u)-\frac{Re}{Fr^2}-Re\, a_y \sin{(t)},\\
&v=\partial_{t}h+u~\partial_{x}h~~~~\mbox{at}~~~~y=h(x,t),\\
&\biggl[\biggl\{1-(\partial_{x}h)^2\biggr\}(\partial_{y}u+\partial_{x}v)-2(\partial_{x}u-\partial_{y}v)\partial_{x}h\biggr]+\mu_H\biggl[(\partial_{x}u-\partial_{y}v)\biggl\{1-(\partial_{x}h)^2\biggr\}+2(\partial_{y}u+\partial_{x}v)\partial_{x}h\biggr]=0 \nonumber\\ &\hspace{11cm} \quad \text{at}\quad y=h(x,t),\\
&p-p_a=\frac{1}{Re\biggl\{1+(\partial_{x}h)^2\biggr\}}\biggl[2\left\{\partial_{x}u\,(\partial_{x}h)^2-(\partial_{y}u+\partial_{x}v)\partial_{x}h+\partial_{y}v\right\}+\mu_H\biggl\{(\partial_{y}u+\partial_{x}v)\biggl\{1-(\partial_{x}h)^2\biggr\}\nonumber\\
&~~~~~~~~~~~~~~~~~~~~~~~~~~~~~~~~~~~~~~~~~~~~~-2(\partial_{x}u-\partial_{y}v)\partial_{x}h\biggr\}\biggr]-We\frac{\partial_{xx}h}{\biggl\{1+(\partial_{x}h)^2\biggr\}^{\frac{3}{2}}}~~~\text{at}~~~~y=h(x,t),\\
&u=0\quad\mbox{and}\quad v=0~~~~\mbox{at}~~~~y=-1.\label{2.16}
\end{align}
The ratio of odd to even viscosity coefficient is symbolized by the new term $\displaystyle \mu_H=\frac{\mu^{o}}{\mu^{e}}$ renamed it as odd or Hall viscosity coefficient, where the Reynolds number $\displaystyle Re=\frac{\rho\omega^2d^2}{\mu^{e}}$ addresses the impact forcing frequency, the Froude number $\displaystyle Fr=\frac{\omega d}{\sqrt{gd}}$ is the gravity effect, and the surface tension effect is denoted by the Weber number $\displaystyle We=\frac{\sigma}{\rho\omega^2d^3}$.

The equations of motion including the boundary conditions for the mean flow with uniform film thickness are given by
\begin{align}
   & \partial_tU_b=\frac{\sin{\alpha}}{Fr^2}+\frac{1}{Re}\partial_{yy}U_b-a_x\sin{t},\label{2.17}\\
   & \partial_yP_b=\frac{\mu_H}{Re}\partial_{yy}U_b-\frac{\cos{\alpha}}{Fr^2}-a_y\sin{(t)},\\
   &\partial_yU_b=0 \quad \text{and}\quad P_b-p_a=\frac{\mu_H}{Re}\,\partial_yU_b~~~~\mbox{at}~~~~y=1,\\
   & U_b=0~~~~\mbox{at}~~~~y=-1\label{2.21},
\end{align}
where $\displaystyle  p_a=\frac{p}{\rho\omega^2H^2}$ stands for the nondimensional ambient pressure.
The above unsteady base flow Eqs.~\eqref{2.17}--\eqref{2.21} for an
unperturbed parallel flow is solved analytically, and its exact solutions take the following form
\begin{align}
&U_b(y,t)=\frac{Re\sin{\alpha}}{2Fr^2}(3+2y-y^2)+a_x\cos{t}-a_x\mathcal{R}\bigg[F(y,t)\bigg]
\end{align}
and
\begin{align}
& P_b(y,t)=p_a+\bigg[\frac{\cos{\alpha}}{Fr^2}+\frac{\mu_H}{Fr^2}\sin{\alpha}+a_y\,\sin{(t)}\bigg](1-y)-\frac{\mu_H}{Re}\,a_x\,\bigg(\mathcal{R}\bigg[J(y,t)\bigg]\bigg),\label{2.24}
\end{align}
where  $\displaystyle F(y,t)=\frac{\cosh{\bigg\{\beta(1+i)(y-1)\bigg\}}}{\cosh{\bigg\{2\beta(1+i)\bigg\}}}\exp{(it)}$, $\displaystyle J(y,t)=\partial_yF(y,t)$, and the real part of the complex function is defined by $\mathcal{R}[\dots]$.  

\section{Time-dependent Orr-Sommerefeld boundary value problem} \label{BVP}
The time-dependent OS BVP is obtained by assuming an infinitesimal perturbation of the unsteady, unidirectional parallel flow solution. A thorough derivation of the time-dependent OS BVP is described by seminal work of \citet{woods1995instability} and \citet{samanta2009effect, samanta2021instability}.
% For the current fluid flow model, one can formulate the following form of the time-dependent OS BVP:
For the current fluid flow model, the time-dependent OS BVP can formulated in the following form:
\begin{align}
&Re\bigg(\partial_t\phi^{''}-k^2\partial_t\phi\bigg)=\bigg(\phi^{''''}-2k^2\phi^{''}+k^4\phi\bigg)-\mathsf{i}\,k\,Re\bigg[U_b~\bigg(\phi^{''}-k^2\phi\bigg)-U_b^{''}\phi\bigg]=0,\label{e25}\\
&\bigg(\phi^{''}+k^2\phi+2\,\mathsf{i}\,k\,\mu_H\,\phi^{'}\bigg)+\bigg(U_b^{''}+2\,\mathsf{i}\,k\,\mu_H\, U_b^{'}\bigg)\eta=0~~~~\mbox{at}~~~~y=1,\\
&Re\,\partial_t\phi^{'}=\biggl[-\mathsf{i}\,k\,Re\biggl\{\frac{\cos{\alpha}}{Fr^2}+a_y\,\sin{(t)}+\frac{\mu_H\sin{\alpha}}{Fr^2}+\frac{a_x\,\mu_H}{Re}\mathcal{R}[\mathcal{D}(t)]\biggr\}-2\,k^2\,U_b^{'}-\mathsf{i}\,k\,\mu_H\,U_b^{''}\nonumber\\
&\quad\quad -\mathsf{i}k^3\,We\,Re\biggr]\eta+\biggl[\phi^{'''}-3k^2\phi^{'}-2\,\mathsf{i}\,k^3\,\mu_H\,\phi\biggr]-\mathsf{i}k\,Re\biggl[U_b\phi^{'}-U_b^{'}\phi\biggr]=0 ~~~~\mbox{at}~~~~y=1,\\
&\partial_t\eta=-\mathsf{i}\,k\,U_b\,\eta-\mathsf{i}\,k\,\phi~~~\mbox{at}~~y=1,\\
&\phi^{'}=0 \quad \text{and} \quad \phi=0~~~~\mbox{at}~~~~y=-1,
\label{e29}
\end{align}
where the prime symbol $^{'}$ stands for the differentiation with respect to the independent variable $y$ and $\displaystyle\mathcal{D}(t)=\frac{2\,\mathsf{i}\,\beta^2}{\cosh{[2\beta(1+\mathsf{i})]}}\exp{(\mathsf{i}\,t)}$. $\phi(y,~t)$ is the amplitude of the perturbation stream function and $\eta(t)$ is the amplitude of the surface deformation. Here $k$ defines the wavenumber in the streamwise direction of an infinitesimal disturbance, where the solution of the infinitesimal disturbance adopts the classical normal mode form $\Tilde{X}(x,y,t)=X(y,t)\exp{(\mathsf{i}\,k\,x)}$, where $X(y,t)$ is the amplitude of the arbitrary perturbed function $\Tilde{X}(x,y,t)$.
\section{Numerical technique}
This section briefly focuses on the numerical procedure to solve the time-dependent OS BVP (Eqs.~\eqref{e25}-\eqref{e29}) associated with the current flow configuration for disturbance of arbitrary wavenumbers. 
In this study, a distinct numerical method, as discussed by \citet{woods1995instability} and \citet{samanta2021instability}, is implemented to resolve the time-dependent OS BVP. Incorporating the Chebyshev spectral collocation method \cite{schmid2001transition}, the time-dependent OS BVP (Eqs.~\eqref{e25}-\eqref{e29}) can be rewritten as
\begin{align}
    &\mathcal{B}\partial_t\Theta=\mathcal{A}\Theta + (N_c\cos{t}+N_s\sin{t})\Theta,\label{e30}
\end{align}
where $A$, $B$, $N_c$, and $N_s$ define square matrices of order $(m+2)\times(m+2)$ and the column matrix $\Theta=\biggl[\phi_0, \phi_1, \phi_2,\dots,\phi_m,\eta\biggr]^{T}$ is of order $(m+2)$ with $m$ Chebyshev modes. Here $T$ marks the transpose of a matrix. The Folquet theory \cite{or1997finite, samanta2017linear, samanta2020effect, samanta2021instability} is employed to solve the Eq.~\eqref{e30}. Subsequently, the time-dependent function $\Theta(t)$ can be expansively represented through a truncated complex Fourier series as 
\begin{align}
    & \Theta(t)=\sum^{n=-\mathcal{K}_t}_{n=\mathcal{K}_t}\Theta_n\exp{[(\mathsf{i}n+\delta)t]}\label{e31}
\end{align}
with constant-coefficient column vectors $\Theta_n$,  the complex Floquet exponent $\delta=\delta_r+\mathsf{i}\delta_i$, and the integers $n$ and $\mathcal{K}_t$.
% with constant-coefficient column vectors $\Theta_n$, $\delta=\delta_r+\mathsf{i}\delta_i$ represents the complex Floquet exponent, and $n$ and $\mathcal{K}_t$ are integers.
Here, the real part $\delta_r$ of the Floquet exponent represents the temporal growth rate of the infinitesimal perturbation. Positive values of $\delta_r$ signify exponential growth of the disturbance amplitude over time, indicating instability in the corresponding flow. Conversely, if the values of $\delta_r$ are negative, the infinitesimal disturbance decays exponentially with time, implying stability in the associated flow. Further, for a given set of parameters, $\delta_r=0$ determines the boundary between the stability and instability zones (i.e., the marginal stability curve).
Note that, in our study, we have emphasized both harmonic
($\delta_i=0$) and subharmonic ($\delta_i=1/2$) solutions instead of the harmonic solution, generally noticeable in the case of fluid flow down a horizontal vibrating plane.
% In our study, we have emphasized both subharmonic ($\delta_i=1/2$) and harmonic ($\delta_i=0$) solutions rather than the harmonic solution merely observed for a liquid flow over a horizontal oscillatory plane. 
% Substituting Eq.~\eqref{e31} in the matrix differential Eq.~\eqref{e30} and collecting the coefficient of $\exp{[(\mathsf{i}n+\delta)t]}$, one can
% obtain a recurrence relation 
Accumulating the coefficients of $\exp{[(\mathsf{i}n+\delta)t]}$ from the resulting equation obtained upon the substitution of Eq.~\eqref{e31} into the Eq.~\eqref{e30}, a matrix difference equation can be obtained as  
\begin{align}
    & \bigg[\mathcal{A}-(\delta+\mathsf{i}r)\mathcal{B}\bigg]\Theta_r+\mathcal{N}\Theta_{r+1}+\mathcal{N}^{*}\Theta_{r-1}=0,\label{e32}
\end{align}
where $\mathcal{N}^{*}$ represents the complex conjugate of $\displaystyle\mathcal{N}=\frac{\mathcal{N}_c+\mathsf{i}\mathcal{N}_s}{2}$. Equation~\eqref{e32} is a linear system with respect to the variables $\Theta_{r+1}$, $\Theta_{r}$ and $\Theta_{r-1}$, where
coefficients refers to the square matrices. The linear system Eq.~\eqref{e32} can be recast as a
generalized matrix eigenvalue problem
\begin{align}
    &\mathcal{F}\mathcal{Y}=\delta\mathcal{G}\mathcal{Y}.\label{e33}
\end{align}
Here, the Floquet exponent $\delta$ is the eigenvalue, $\mathcal{Y}=\bigg[\Theta_{-\mathcal{K}_t},\Theta_{-(\mathcal{K}_t-1)},\Theta_{-(\mathcal{K}_t-2)},\dots,\Theta_{\mathcal{K}_t-2},\Theta_{\mathcal{K}_t-1},\Theta_{\mathcal{K}_t}\bigg]$ is the column matrix, $\mathcal{F}$ is the block tri-diagonal matrix, and $\mathcal{G}$ is the block diagonal matrix. The forms of $\mathcal{F}$ and $\mathcal{G}$ are mentioned in the Appendix \ref{mat}. 
% The positive real part of the Floquet exponent (i.e., $\delta_r>0$) indicates the linear instability of the flow system, whereas $\delta_r<0$ corresponds to the stability of the flow.   
% The fluid flow will be linearly stable if the real parts of the Floquet exponent are positive (i.e., $\delta_r>0$), whereas $\delta_r<0$ corresponds to the stability of the flow system. 

\section{Numerical results} \label{NR}
\begin{figure}[ht!]
\begin{center} 
\subfigure[]{\includegraphics*[width=8cm]{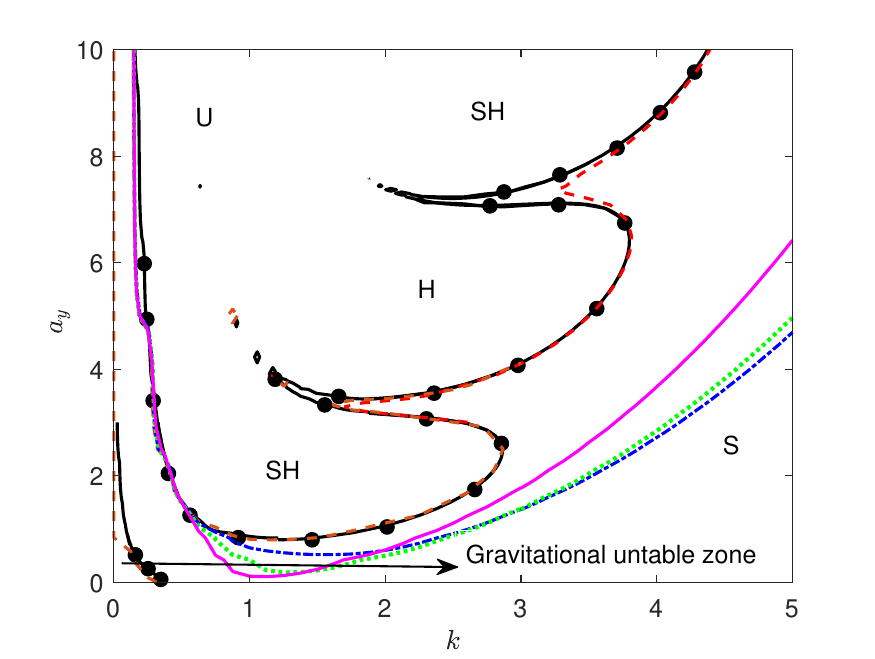}}
\subfigure[]{\includegraphics*[width=8cm]{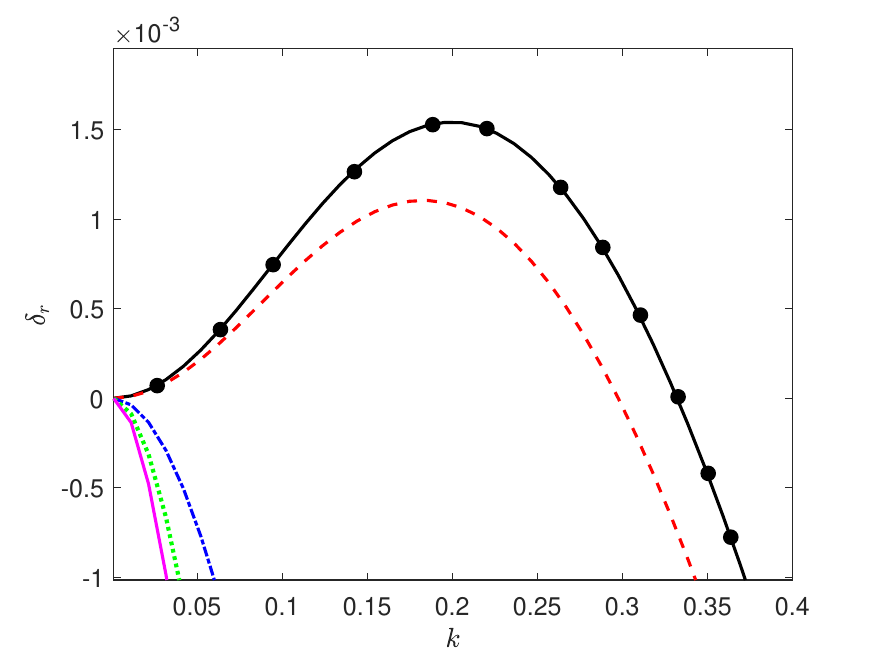}}
 \end{center}
\caption{ Variation of (a) the instability boundary lines in the ($k,a_x$)-plane and (b) the corresponding growth rate $\delta_r$ related to the gravitational instability with $a_y=0.1$ when the odd-viscosity coefficient $\mu_H$ varies. Black, red, blue, green, and magenta lines are the results for $\mu_H=0$, $\mu_H=0.05$, $\mu_H=1$, $\mu_H=2$, and $\mu_H=3$, respectively. Remaining parameter values are $a_x=0$,  $We=0.016$, $Fr^2 = 100$, $Re = 5$, and $\alpha=90^{\circ}$. The red-filled circles are the results of \citet{woods1995instability}.}
\label{fig2}

\end{figure}

In this section, we have discussed the dynamics and instability mechanisms of different existing unstable modes in the current flow problem when the reversal symmetry of time breaks. 
% Prior to delving into the results, it is imperative to verify the numerical code by comparing it with established findings. 
In Fig.~\ref{fig2}(a), the marginal stability curve in the ($k,~a_x$)-plane is displayed when $\mu_H=0$, $a_y=0.1$, $We=0.016$, $Fr^2 = 100$, $Re = 5$, and $\alpha=90^{\circ}$. The symbols 'U' and 'S' are incorporated to define the unstable and stable zones, respectively. Here, only the normal oscillation of the bottom wall is considered (i.e., $a_y\neq0$ and $a_x=0$).
The marginal curve well matches the result presented by \citet{woods1995instability} when the negligible odd-viscosity coefficient $\mu_H$ is taken into account. Fig.~\ref{fig2}(a) illustrates the emergence of three distinct unstable zones delineating gravitational, subharmonic (SH), and harmonic (H) instability when $\mu_H=0$. Gravitational instability manifests in the longwave range, whereas SH and H unstable zones occur within the finite wavenumber range (\cite{woods1995instability}, \cite{samanta2021instability}). The tongue-shaped patterns of unstable SH and H regions are identified in the wide range of unstable wavenumbers, where the first H tongue is higher compared to the first SH tongue. Initially, the SH instability is responsible for the generation of resonant waves. Thus, in the absence of broken time-reversal symmetry (i.e., $\mu_H=0$), the change in the forcing amplitude in cross-stream oscillation resonates SH and H instabilities in different unstable regions of wavenumber. The stable region between gravitational and SH instabilities is enhanced for the presence of the odd viscosity $\mu_H$ in the fluid flow. Consequently, the inclusion of the odd viscosity coefficient $\mu_H$ decelerates the transition process from gravitational to SH instability. Further, the odd viscosity $\mu_H$ attenuates the stable zone between the SH and H instability emerging in the finite wavenumber range, and this stable zone entirely vanishes for $\mu_H\geq 1$. This fact fastens the transition from SH to H instability. 
% Consequently, the inclusion of the odd viscosity coefficient $\mu_H$ accelerates the transition process from gravitational to SH instability.
Moreover, for a low forcing amplitude of the cross-stream oscillation, the corresponding growth rate result related to the gravitational instability, shown in Fig.~\ref{fig2}(b), displays a slowdown of the maximum growth rate as $\mu_H$ goes up. As a result, a higher odd viscosity $\mu_H$ possesses the capacity to delay the transition from gravitational instability to turbulence.

\begin{figure}[ht!]
\begin{center} 
\subfigure[]{\includegraphics*[width=8cm]{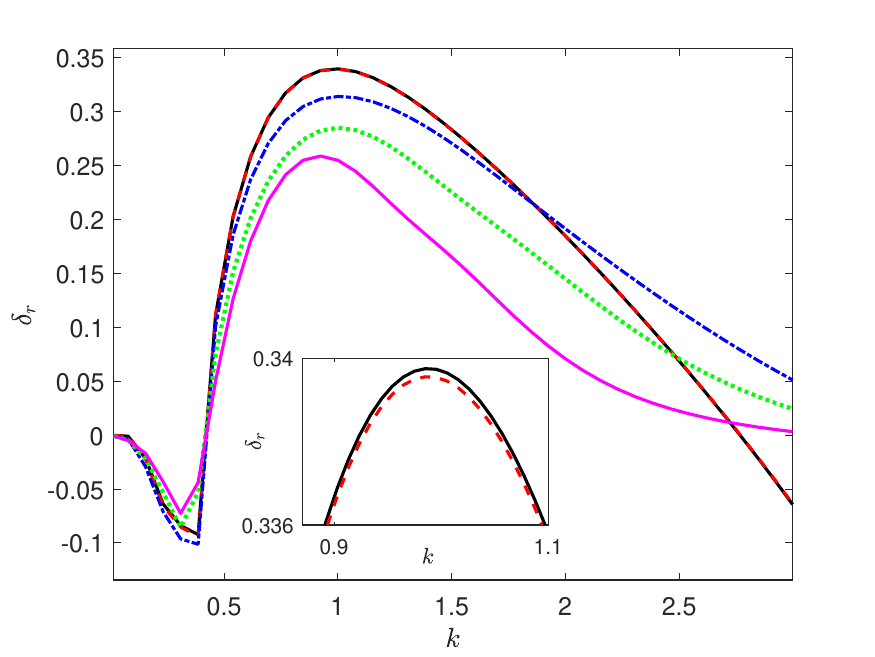}}
\subfigure[]{\includegraphics*[width=8cm]{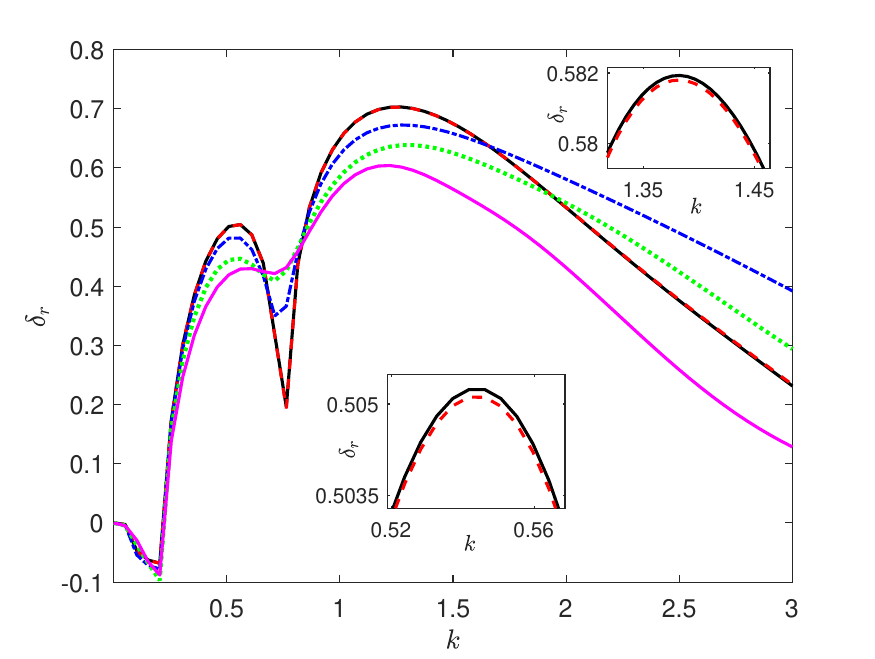}}
 \end{center}
\caption{ Variation of the temporal growth rate $\delta_r$ with  varying $\mu_H$ when (a) $a_y=2$ and (b) $a_y=6$. Black, red, blue, green, and magenta lines are the results for $\mu_H=0$, $\mu_H=0.05$, $\mu_H=1$, $\mu_H=2$, and $\mu_H=3$, respectively. Remaining parameter values are $a_x=0$,  $We=0.016$, $Fr^2 = 100$, $Re = 5$, and $\alpha=90^{\circ}$.
}
\label{fig333}
\end{figure}

% \begin{figure}[ht!]
% \begin{center} 
% \subfigure[]{\includegraphics*[width=8cm]{Fig_6_b}}
% \subfigure[]{\includegraphics*[width=8cm]{Fig_6_d}}
%  \end{center}
% \caption{ Variation of the temporal growth rate $\delta_r$ as a function of wavenumber $k$ with the odd-viscosity coefficient $\mu_H$ varies when (a) $a_y=2$ and (b) $a_y=6$. Black, red, blue, and green lines are the results for $\mu_H=0$, $\mu_H=1$, $\mu_H=2$, and $\mu_H=3$, respectively. The other flow parameters are $a_x=0$,  $We=0.016$, $Fr^2 = 100$, $Re = 5$, and $\alpha=90^{\circ}$.
% }
% \label{fig333}
% \end{figure}

\begin{figure}[ht!]
\begin{center} 
\includegraphics*[width=16cm,height=10cm]{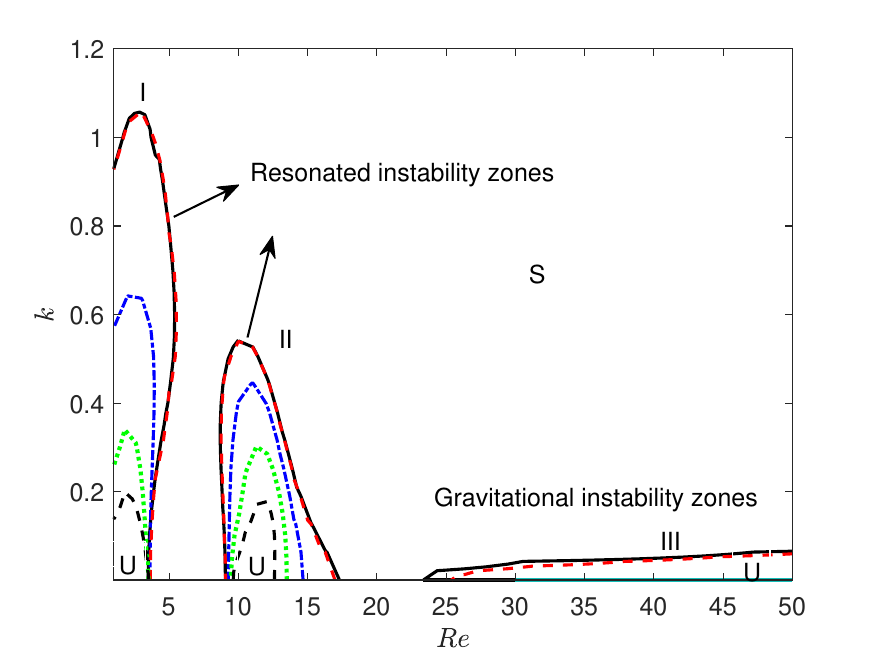}
\end{center}
\caption{ Variation of (a) the instability boundary curves in the (Re-k)-plane with varying $\mu_H$. Black, red, blue, green, and magenta lines are the results for $\mu_H=0$, $\mu_H=0.05$, $\mu_H=1$, $\mu_H=2$, and $\mu_H=3$, respectively. Remaining parameter values are $a_x=6$, $a_y=0$, $We=0.016$, $Fr = 100$, and $\alpha=90^{\circ}$. }
\label{fig22}
\end{figure}

\begin{figure}[ht!]
\begin{center} 
\subfigure[]{\includegraphics*[width=5.4cm]{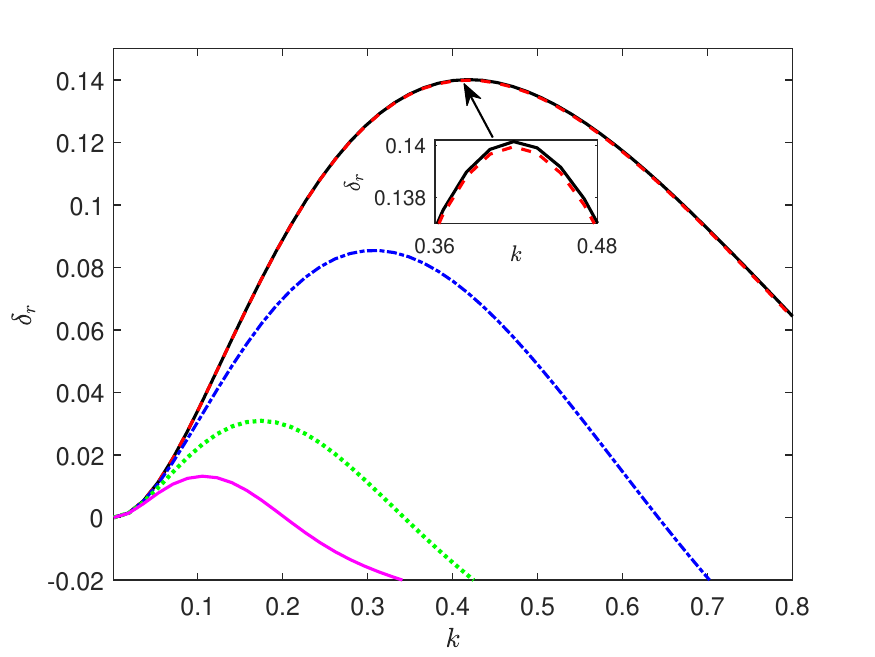}}
\subfigure[]{\includegraphics*[width=5.4cm]{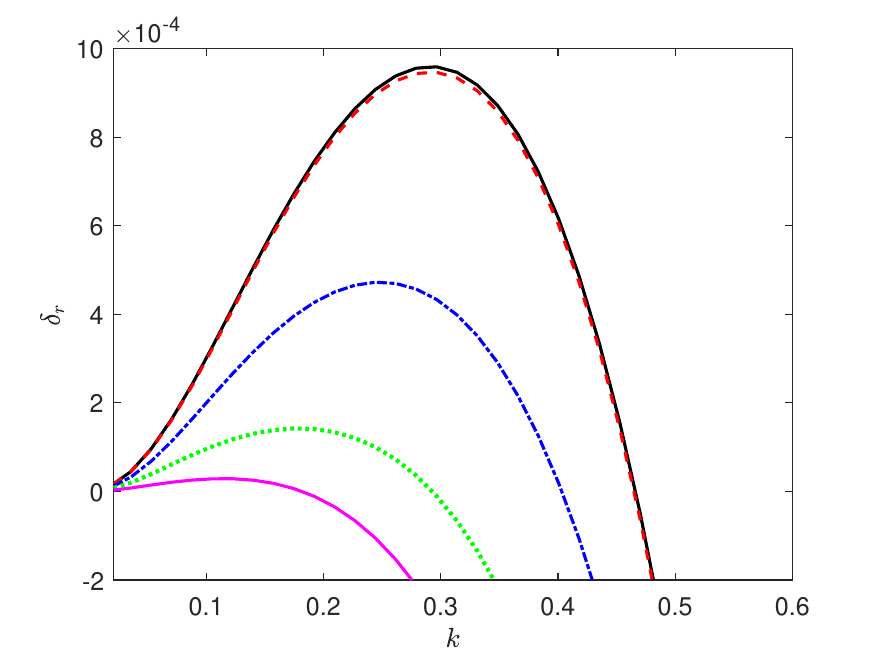}}
\subfigure[]{\includegraphics*[width=5.4cm]{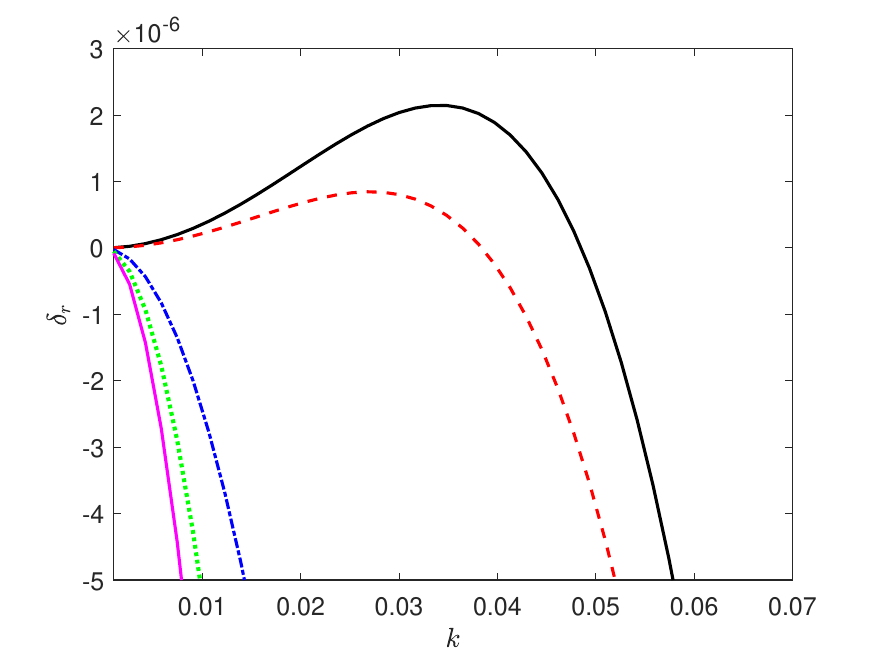}}
 \end{center}
\caption{Variation of the temporal growth rate $\delta_r$ with varying $\mu_H$ when (a) $Re=2$, (b) $Re=12$, and (c) $Re=35$. Black, red, blue, green, and magenta lines are the results for $\mu_H=0$, $\mu_H=0.05$, $\mu_H=1$, $\mu_H=2$, and $\mu_H=3$, respectively. Remaining parameter values are $a_x=6$, $a_y=0$, $We=0.016$, $Fr = 100$, and $\alpha=90^{\circ}$.
}
\label{fig55}

\end{figure}

\begin{figure}[ht!]
\begin{center} 
\subfigure[$a_x\neq0$ and $a_y=0$]{\includegraphics*[width=8cm]{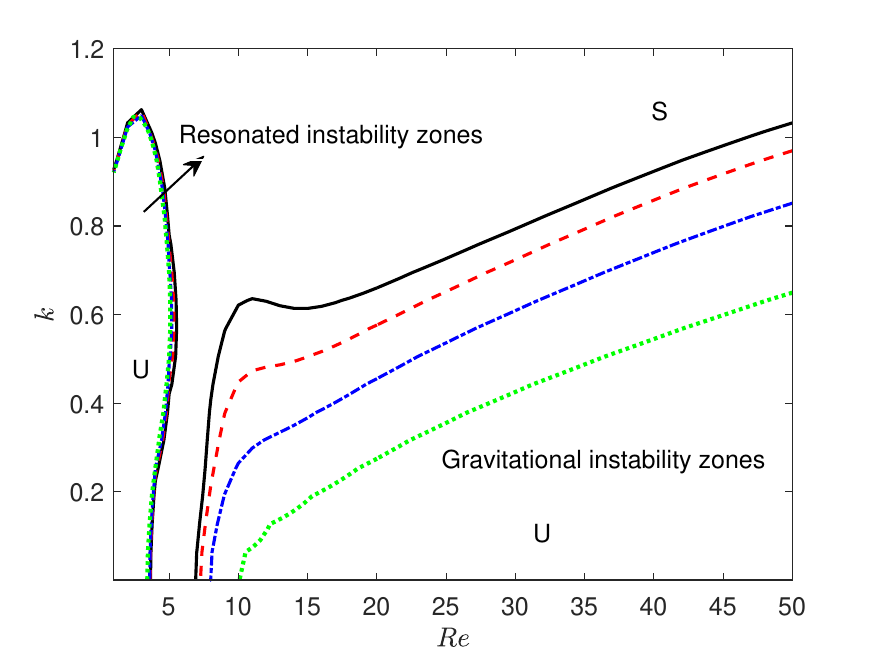}}
\subfigure[$a_x=0$ and $a_y\neq0$]{\includegraphics*[width=8cm]{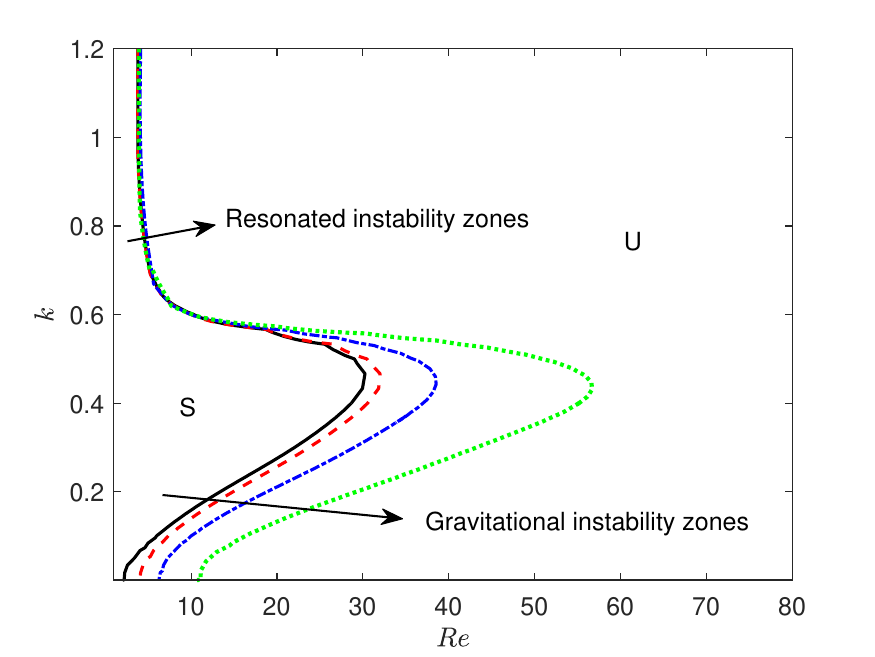}}
 \end{center}
\caption{ Variation of the instability boundary lines in the $(Re-k)-$ plane with varying $\alpha$ when (a) $a_x=6$ and $a_y=0$, and (b) $a_x=0$ and $a_y=1$. Black, red, blue, and green lines are the results for $\alpha=90^{\circ}$, $\alpha=70^{\circ}$, $\alpha=50^{\circ}$, and $\alpha=30^{\circ}$, respectively. Remaining parameter values are $\mu_H=0.05$, $We=0.016$ and $Fr^2 = 100$. 
}
\label{fig56666}

\end{figure}

Also, to detect the behavior of Faraday waves, the temporal growth rate ($\delta_r$) associated with Fig.~\ref{fig2}(a), is demonstrated with different forcing amplitudes $a_y$ (see, Fig.~\ref{fig333}(a) for $a_y=1$ and Fig.~\ref{fig333}(b) for $a_y=6$). When $a_y=1$, the growth rate, as in Fig.~\ref{fig333}(a), displays the sole hump related to SH instability emerging in the finite wavenumber zone. 
% The growth rate of subharmonic instability is much higher than that of gravitational instability.
The odd viscosity coefficient $\mu_H$ decreases the maximum growth rate of SH instability, thereby exerting a stabilizing effect.
% The effect of $\mu_H$ on the gravitational instability is very weak compared to the subharmonic instability. 
On the other hand, upon increasing the forcing amplitude to $a_y=6$, the temporal growth profile unveils a new hump related to H instability (see, Fig.~\ref{fig333}(b)), including SH instability. The wavenumber of H instability is much greater than the wavenumber of the first SH instability, whereas H instability is much stronger than SH instability. The higher odd viscosity lowers the maximum growth rate of both SH and H instabilities and promotes the stabilizing impact on those instabilities.
% Here, subharmonic instability is located in the long-wave regime, whereas subharmonic and harmonic instabilities occur in the finite-wavelength zone.  Moreover, it is noticed that the maximum growth rate of both harmonic and subharmonic instabilities weaken as long as $\mu_H$ increases.

% , but compared to the harmonic instability, $\mu_H$ has a weak impact on the gravitational and subharmonic instabilities.

In Fig.~\ref{fig22}, the neural stability curves in the ($Re,~k$) plane with different odd viscosity parameters $\mu_H$ are demonstrated when the bottom wall solely oscillates in the streamwise direction (i.e., $a_x\neq0$ and $a_y=0$). When $\mu_H=0$ (i.e, the time-reversal symmetry does not break), three different unstable regions emerge that are separated by the stable range of Reynolds number, where the first two unstable regions I and II are generated because of the bottom wall's streamwise oscillation, and the unstable region III is instigated by the gravity driving force. 
That means the gravitational instability is not only responsible for driving the fluid flow's primary instability but also the resonated instability caused by the streamwise oscillation of the bottom wall. 
% That means the primary instability of the fluid flow is not only driven by gravitational instability but also resonated instability due to the streamwise oscillation of the bounding plane. 
Another important fact is that the higher values of odd viscosity enhance the stable zone of Reynolds number generated among the separated unstable zones I, II, and III. Consequently, the higher odd viscosity slows the transition process from resonance to gravitational instabilities. Further, at lower $Re$ values, the resonated unstable region dominates the gravitational instability region. The unstable zones associated with the resonance and gravity waves shrink with the increasing values of $\mu_H$, leading to a stabilizing influence on these unstable zones. 

% one can stabilize those instabilities. 

% get a continuous transition from the resonated instability to gravitational
% instability. Notably, the gravitational instability zone expands for a higher value of $\mu_H$.

In order to clarify the above fact,  the associated growth rate curves are displayed in Fig.~\ref{fig55} related to the resonated unstable regions I (see, Fig.~\ref{fig55}(a) for $Re=2$) and II (see, Fig.~\ref{fig55}(b) for $Re=12$) and for the unstable region III (see Fig.~\ref{fig55}(c) for $Re=35$). The values of $Re$ are chosen from the unstable region, as depicted in Fig.~\ref{fig22}. It is identified that the growth rates associated with resonated instabilities I, as in Fig.~\ref{fig55}(a), and II, as in Fig.~\ref{fig55}(b) in the finite wave number range diminishes for higher odd viscosity $\mu_H$. However, for a sufficiently higher value $Re=35$, the gravitational instability arises in the longwave region, and the corresponding growth rate attenuates as soon as $\mu_H$ increases. These outcomes entirely agree with the result of Fig.~\ref{fig22}. Thus, the odd viscosity has the potential to stabilize the resonated instability along with the gravitational instability generated in the falling liquid film over an oscillating plane when the time-reversal symmetry breaks.  

% \textcolor{blue}{an additional hump in the longwave region depicting the gravitational instability does not exist, whereas the higher odd viscosity coefficient intensifies the growth rate (see Fig.\ref{fig55}(a)) of the existing resonated unstable regions I and II. }

% \textcolor{red}{need to modify the physics of Fig.5}
% However, for sufficiently higher $Re=30$, gravitational instability region III emerges in the longwave region and the associated growth rate (see Fig.\ref{fig55}(a)) enhances with the higher value of odd viscosity. 
    
\begin{figure}[ht!]
\begin{center} 
\includegraphics*[width=16cm,height=10cm]{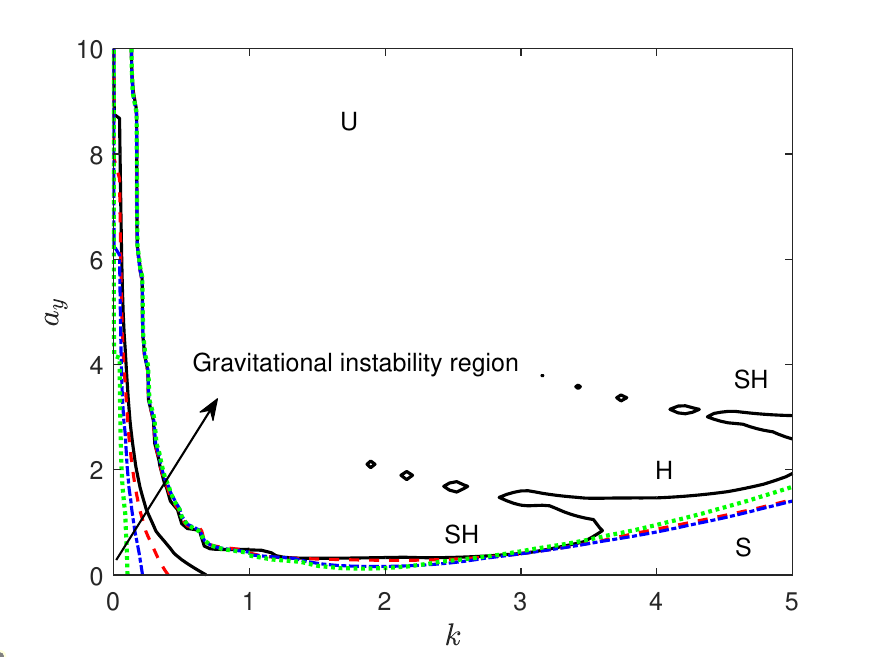} {Continuation of Fig.~\ref{fig2}}
\end{center}
\caption{ Variation of (a) the instability boundary lines in the $(a_y-k)-$ plane with varying $\mu_H$. Black, red, blue, and green lines are the results for $\mu_H=0$, $\mu_H=1$, $\mu_H=2$, and $\mu_H=3$, respectively. Remaining parameter values are $Re = 20$, $a_x=0$, $We=0.016$, $Fr^2 = 100$, and $\alpha=90^{\circ}$. }
\label{fig7}
\end{figure}

Also, we have tracked out the effect of the angle of inclination $\alpha$ of the bounding wall on the resonated instability owing to the oscillation of the bottom substrate and gravitational instabilities. Figure~\ref{fig56666} represents the neutral stability curves in $Re-k$ plane for different inclination angles $\alpha$ when both cases (i) lateral (see, Fig.~\ref{fig56666}(a)) and (ii) normal (see, Fig.~\ref{fig56666}(b)) oscillations are individually considered.  As in Fig.~\ref{fig56666}(a), When $a_x=6$ and $a_y=0$ (i.e., only the streamwise oscillation of the bounding wall), no noticeable deviation in the resonance curves is observed with the varying angle $\alpha$, whereas the gravitational instability emerging in the longwave region drastically shrinks, followed by the increasing of the critical Reynolds number. On the other hand, for $a_x=0$ and $a_y=1$ (i.e., only the normal oscillation of the bounding wall), the inclination angle $\alpha$ has negligible impact on the instability regions corresponding to the resonance waves (see, Fig.~\ref{fig56666}(b)), while, as expected, the gravitational instability regions significantly attenuate as soon as the values of $\alpha$ reduce. Thus, the influence of the angle of inclination is only perceptible for gravitational instability, whereas parametric resonances experience a very weak impact.

Figure~\ref{fig7} displays the continuation of Fig.~\ref{fig2} for sufficiently increasing the Reynolds number $Re=20$. The stability boundary line demonstrates that the resonated waves form a tongue-like pattern in the absence of broken-time reversal symmetry. It is observed that gravitational instability increases for sufficiently high inertia force. Also, the unstable zone of the SH and H instabilities enhances for higher value $Re=20$ by reducing the critical forcing amplitudes for the first SH and first H instabilities. Note that, when the time-reversal symmetry breaks (i.e., $\mu_H\neq 0$) the stable region of sufficiently higher wavenumbers between SH and H instabilities vanishes, making the transition process from SH to H instability faster. Physically, one can produce SH and H
instabilities, comparatively, at lower forcing amplitudes $a_y$ for a cross-stream oscillatory
odd-viscosity-induced fluid flow. 
Moreover, a higher value of odd viscosity reduces the unstable zone of gravitational instability by dissipating gravity waves. Consequently, higher odd viscosity enhances the stable range between gravitational and SH instabilities and weakens the transition process from gravitational instability to SH instability.

\begin{figure}[ht!]
\begin{center} 
\subfigure[]{\includegraphics*[width=8cm]{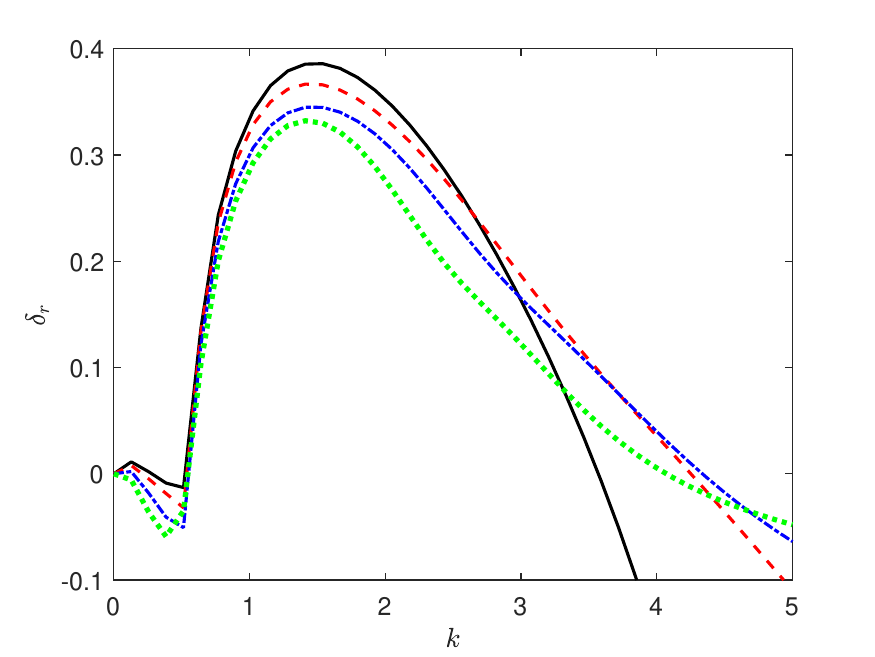}}
\subfigure[]{\includegraphics*[width=8cm]{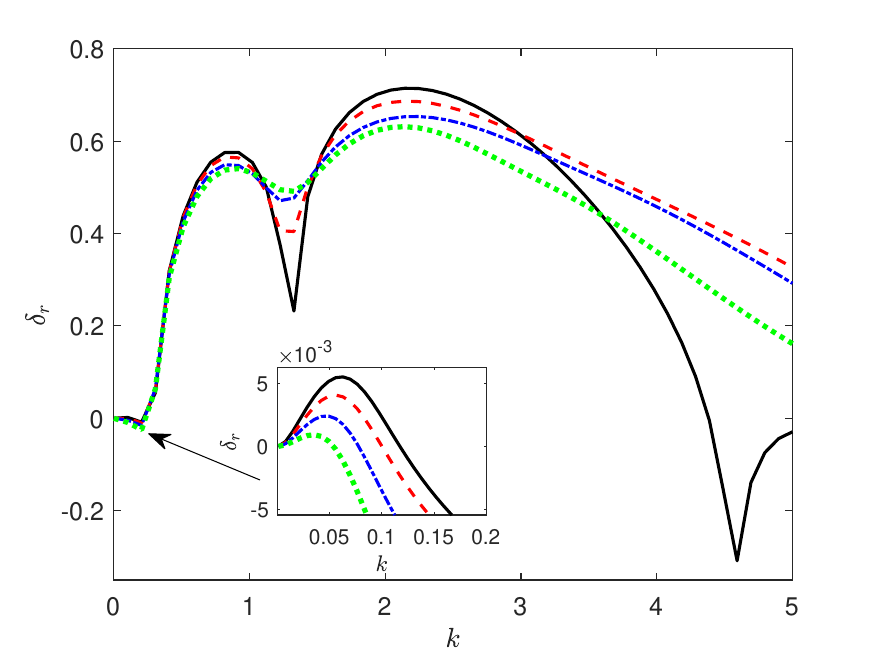}}
 \end{center}
\caption{ Variation of the temporal growth rate $\delta_r$ with varying $\mu_H$ when (a) $a_y=1$ and (b) $a_y=3$ with $Re=20$. Black, red, blue, and green lines are the results for $\mu_H=0$, $\mu_H=1$, $\mu_H=2$, and $\mu_H=3$, respectively. Remaining parameter values are $a_x=0$,  $We=0.016$, $Fr^2 = 100$, and $\alpha=90^{\circ}$.
}
\label{fig352}
\end{figure}

The corresponding temporal growth rate curves with different $\mu_H$ are plotted in Fig.~\ref{fig352} when the forcing amplitude $a_y=1$, as in Fig.~\ref{fig352}(a) and $a_y=3$, as in Fig.~\ref{fig352}(b). Fig.~\ref{fig352}(a) depicts two humps, where the growth rate of both gravitational and SH instabilities decrease as long as odd viscosity increases. This fact assures that a higher value of odd viscosity can weaken both gravitational and SH instabilities. On the other side, for a comparatively higher forcing amplitude of the cross-stream oscillation ($a_y=3$), the growth rate curve displays three humps instead of two humps, which is fully consistent with the neutral curve results in Fig.~\ref{fig7}. The new hump in the finite wavenumber region corresponds to the H instability, which diminishes with an increasing value of odd viscosity $\mu_H$.

\begin{figure}[ht!]
\begin{center} 
\includegraphics*[width=16cm,height=10cm]{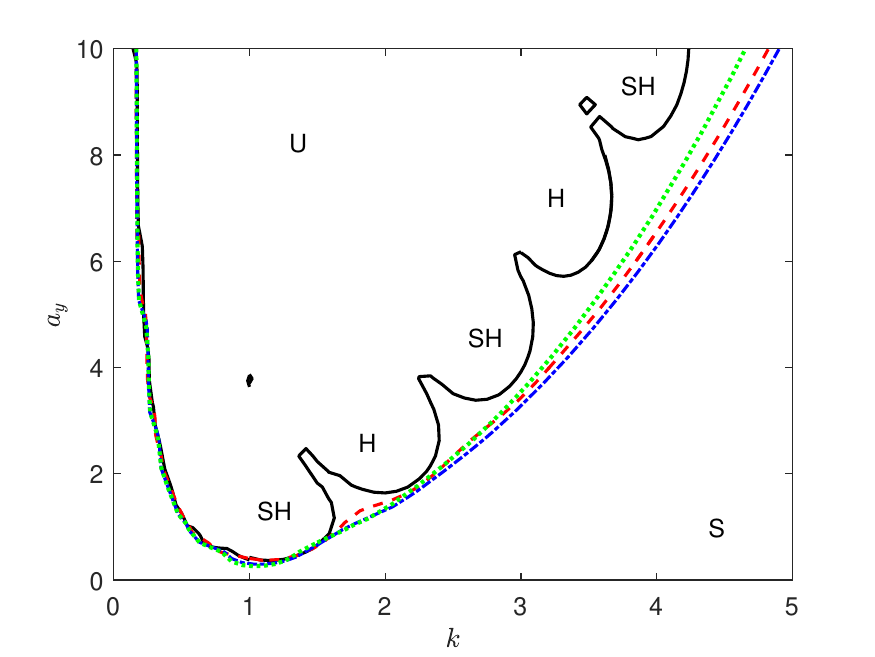}
\end{center}
\caption{ Variation of the instability boundary lines in the $(a_y-k)-$ plane with varying $\mu_H$. Black, red, blue, and green lines are the results for $\mu_H=0$, $\mu_H=1$, $\mu_H=2$, and $\mu_H=3$, respectively. Remaining parameter values are $Re = 20$, $a_x=0$, $We=0.16$, $Fr^2 = 100$, and $\alpha=1^{\circ}$. }
\label{fig77}
\end{figure}

\begin{figure}[ht!]
\begin{center} 
\subfigure[]{\includegraphics*[width=8cm]{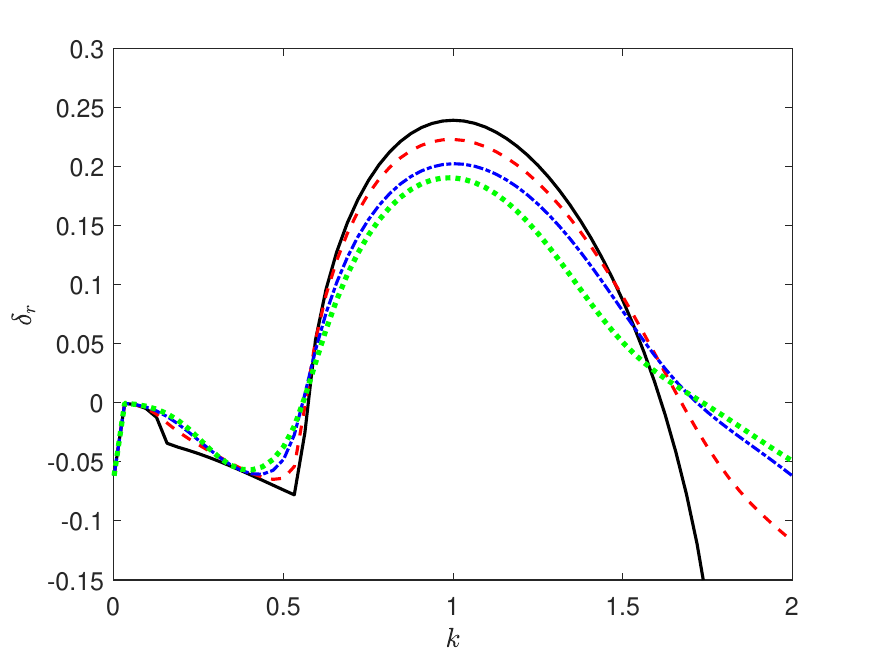}}
\subfigure[]{\includegraphics*[width=8cm]{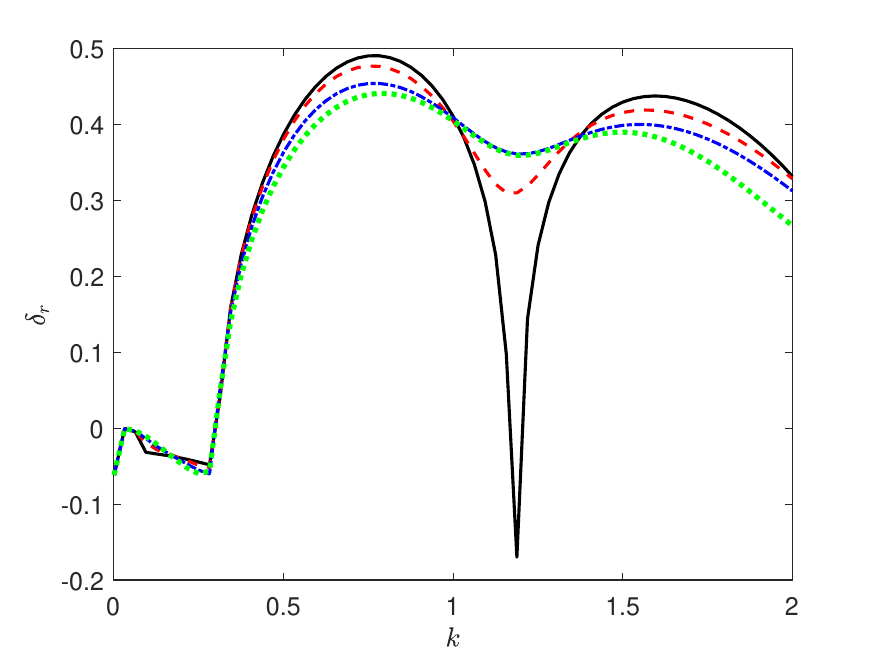}}
 \end{center}
\caption{Variation of the temporal growth rate $\delta_r$ with varying $\mu_H$ when (a) $a_y=1$ and (b) $a_y=3$. Black, red, blue, and green lines are the results for $\mu_H=0$, $\mu_H=1$, $\mu_H=2$, and $\mu_H=3$, respectively. Remaining parameter values are $a_x=0$, $We=5$, $Re=10$, $Fr^2 = 1$, and $\alpha=1^{\circ}$.
}
\label{fig88}
\end{figure}
Now, the neutral stability curve for different odd viscosity is demonstrated in Fig.~\ref{fig77} to understand the instability behavior of Faraday waves when the liquid layer is comparatively thick with high surface tension and low inclination angle. The SH and H unstable regions are displayed alternatively in this tongue-like unstable region when the odd viscosity coefficient is absent. It is to be noted that, the neutral stability curve does not occur in the gravitational unstable region. The reason is that the Reynolds number belongs to the stable zone. Once, the odd viscosity is introduced in the liquid film, the stable zones in the finite wavenumber range between the SH and H instability fully disappear. The critical forcing amplitudes for the SH and H tongues
attenuate for a higher value of odd viscosity. However, the critical forcing amplitude for the first SH tongue always remains lower than that for the first H
tongue. Therefore, introducing lower forcing amplitudes makes it possible to create SH and H resonances in the odd-viscosity-induced thinner-layered liquid with high surface tension flowing
down a slightly inclined oscillating plane.

% Therefore, by introducing lower forcing amplitudes, it is possible to create SH and H resonances in the odd-viscosity-induced thinner-layered liquid with high surface tension flowing
% down a slightly inclined plane. 
% Therefore, it is possible to generate subharmonic and harmonic resonances at lower forcing
% amplitudes for an odd-viscosity-induced thinner-layered liquid with high surface tension flowing
% down a slightly inclined plane.

To make this result portrayed in Fig.~\ref{fig77} consistent, the associated temporal growth rate result is plotted in Fig.~\ref{fig88} for different odd viscosity parameters with various forcing amplitudes of the cross-stream oscillation of the bounding wall. As expected, for lower forcing amplitude $a_y=1$ of cross-stream oscillation, gravitational instability is absent in the temporal growth rate profile (see, Fig.~\ref{fig88}(a)), whereas the SH instability appears and reduces with higher odd viscosity. 
% instability is absent. For lower forcing amplitude $a_y=1$ of cross-stream oscillation, the SH instability appears (see, Fig.~\ref{fig88}(a)) and it reduces with higher odd viscosity. 
However, for comparatively higher forcing amplitude $a_y=3$, both SH and H instabilities arise (see, Fig.~\ref{fig88}(a)) in the wide range of finite wavenumbers, and both types of instability diminish with the increasing value of odd viscosity. Therefore, the resonated waves generated at the liquid surface can be weakened owing to the higher odd viscosity value. These facts totally validate the findings pertained in Fig.~\ref{fig77}.

 \begin{figure}[ht!]
\begin{center} 
\subfigure[]{\includegraphics*[width=8cm]{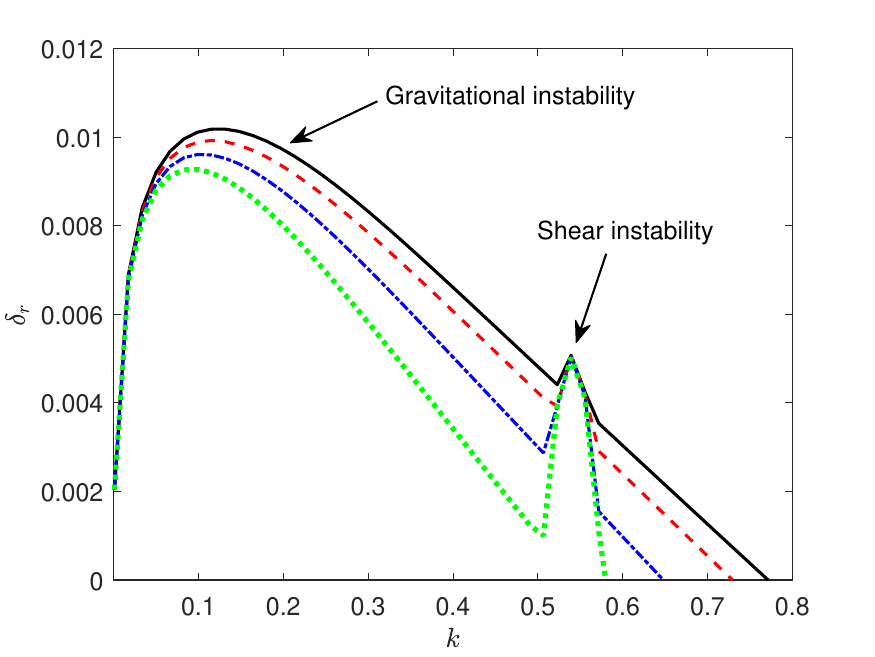}}
\subfigure[]{\includegraphics*[width=8cm]{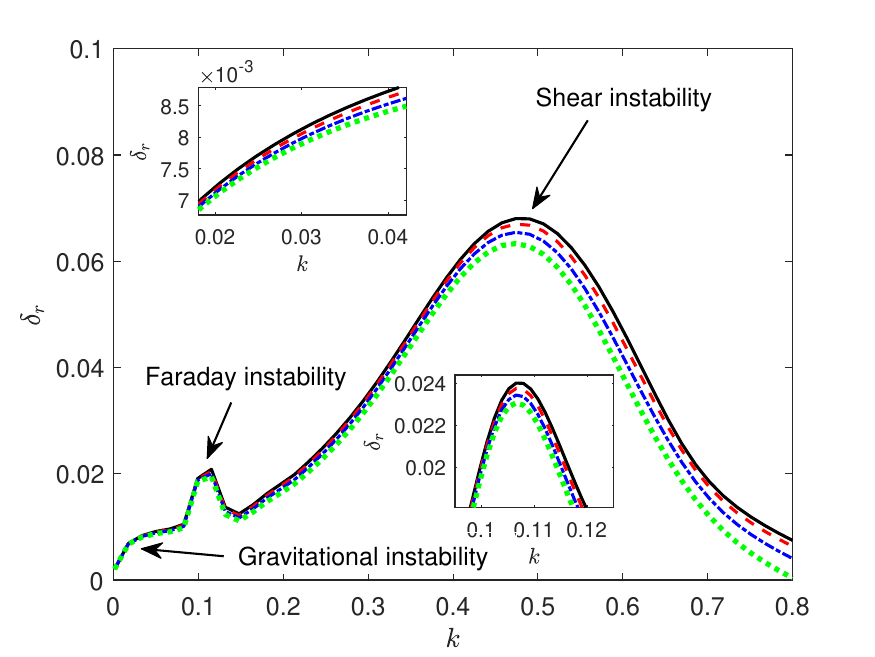}}
 \end{center}
\caption{Variation of the temporal growth rate $\delta_r$ with varying $\mu_H$ when (a) $a_y=0$ and (b) $a_y=3$ with $Re=290$. Black, red, blue, and green lines are the results for $\mu_H=0$, $\mu_H=10$, $\mu_H=20$, and $\mu_H=30$, respectively. Remaining parameter values are $a_x=0$, $We=5$, $Fr^2 = 1$, and $\alpha=1^{\circ}$.
}
\label{fig888}
\end{figure}
To detect the behavior of Faraday waves with different odd viscosity coefficients $\mu_H$ for falling thin film under high inertia force, with high surface tension flowing down a slightly inclined plane, Fig.~\ref{fig888} is plotted by choosing $We = 5$, $Fr=1$, $Re = 290$, and $\alpha=1^{\circ}$ (\citet{woods1995instability}). The novelty that emerges from Fig.~\ref{fig888} is that there are two humps in the temporal growth rate result with no external oscillatory forcing (see, Fig.~\ref{fig888}(a) for $a_y=0$): one is related to gravitational instability in the longwave region, and the other is shear instability in the finite wavenumber range. Thus, the shear mode instability at the liquid surface is observed when the Reynolds number is chosen sufficiently high with a small inclination angle of the bottom wall. The emergence of shear instability is owing to the Reynolds number exceeding its critical level. Here, as $\mu_H$ increases, the growth rate decreases for the gravity and shear waves. On the other hand, as long as the external cross-stream oscillation is incorporated (see, Fig.~\ref{fig888}(b) for $a_y=3$), both SH and H resonated instability arise in the temporal growth along with the gravitational and shear instabilities.  A higher value of odd viscosity weakens all these different types of instability emerging in the flow system by attenuating the corresponding growth rates.

\begin{figure}[ht!]
\begin{center} 
\includegraphics*[width=16cm,height=10cm]{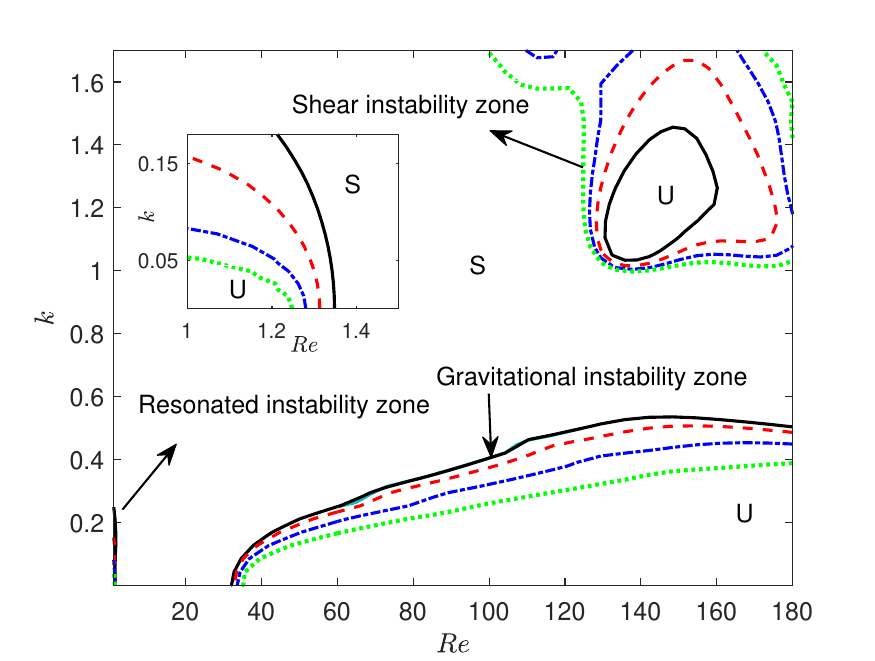}
\end{center}
\caption{ Variation of the instability boundary lines in the $(Re-k)-$ plane with varying $\mu_H$. Black, red, and blue lines are the results for $\mu_H=0$, $\mu_H=1$, $\mu_H=2$, and $\mu_H=3$, respectively. Remaining parameter values are $a_x=4$, $a_y = 0$, $We=5$, $Fr^2 = 1$, and $\alpha=1^{\circ}$. }
\label{fig20}
\end{figure}

 \begin{figure}[ht!]
\begin{center} 
\subfigure[]{\includegraphics*[width=8cm]{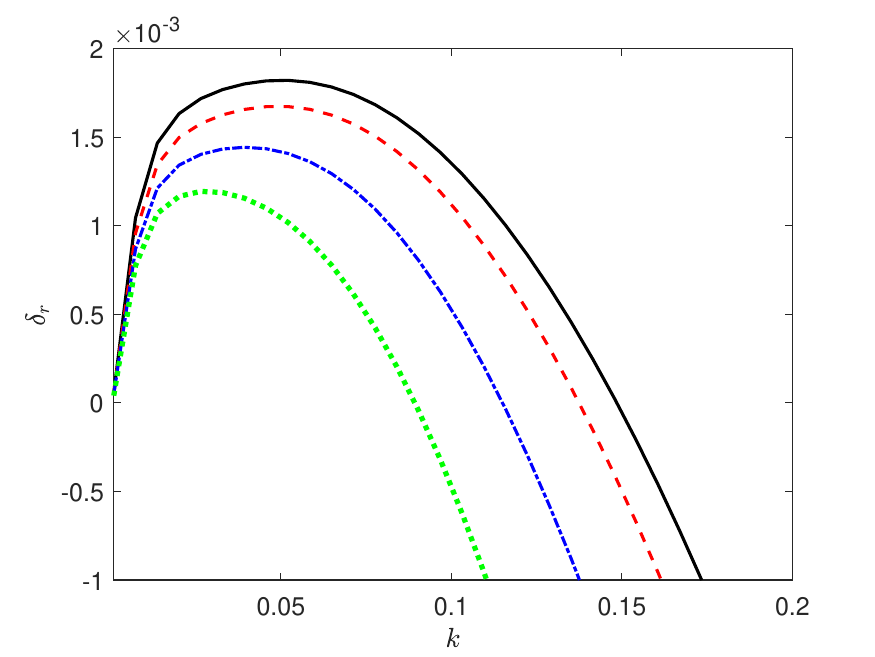}}
\subfigure[]{\includegraphics*[width=8cm]{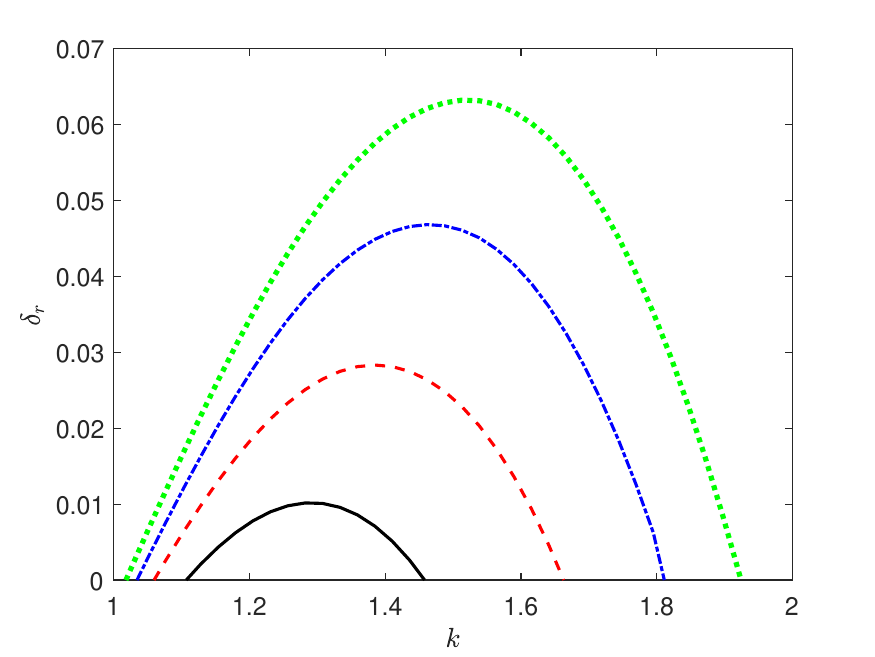}}
 \end{center}
\caption{Variation of the temporal growth rate $\delta_r$ with varying $\mu_H$ when (a) $Re=40$ and (b) $Re=150$. Black, red, blue, and green lines are the results for $\mu_H=0$, $\mu_H=1$, $\mu_H=2$, and $\mu_H=3$, respectively. Remaining parameter values are $a_x=4$, $a_y=0$, $We=5$, $Fr^2 = 1$, and $\alpha=1^{\circ}$.
}
\label{fig2222}
\end{figure}

Now the stability boundary lines with different odd viscosity $\mu_H$ are demonstrated in Fig.~\ref{fig20} when a thin liquid layer with high surface tension flows down over a cross-streamwise oscillatory slightly inclined substrate. The marginal stability curve shows three different unstable zones gravitational, resonated, and shear instabilities. Higher odd viscosity shrinks the unstable longwave zones of gravitational instability by enhancing the critical Reynolds number. Contrapositively, the shear instability existing in the higher wavenumber range intensifies with the increasing value of the odd viscosity coefficient followed by a successive increase of corresponding unstable zones. However, compared to the gravity and shear instabilities, the odd viscosity has a weak but stabilizing influence on the resonated instability. Thus, the surface wave instability generated by the gravity driving force and the oscillating plane can be weakened by considering higher values of odd viscosity. Next, the associated temporal growth rate curves are shown in Fig.~\ref{fig2222} pertaining to the change in gravitational instability when $Re=40$ (see, Fig.~\ref{fig2222}(a)) and shear instability when $Re=150$ (see, Fig.~\ref{fig2222}(b)). A decrease in the growth rate, as in Fig.~\ref{fig2222}(a), of the gravitational instability is noticed with higher values of $\mu_H$. This showcases the stabilizing influence of odd viscosity on the gravity-driving instability waves. On the other hand, the higher odd viscosity amplifies the growth rate of the shear instability, as depicted in Fig.~\ref{fig2222}, manifesting in the higher wavenumber range.

\section{Conclusions} \label{CON}
The instability behavior of falling liquid over an inclined plane that is compelled to vibrate in both the streamwise and cross-stream directions with identical frequency is examined when the time-reversal symmetry breaks. The linear stability analysis is performed for arbitrary wavenumbers by numerically solving the corresponding time-dependent OS BVP, including the Floquet theory.
The numerical outcomes manifest the occurrence of three different unstable regions: gravitational, SH, and H. In general, gravitational instability emerges in the long-wave range, whereas SH and H instabilities exist in the finite wavenumber regime. In comparison to the generation of H instability, the critical forcing amplitude for the generation of SH instability is smaller. This assures the dominance of SH instability in the finite wavenumber region. When the time-reversal symmetry breaks, i.e., the odd viscosity is introduced to the flow system, a higher value of odd viscosity attenuates the stable region between gravitational and SH instabilities in the longwave range. Also, the stable range located in the finite wavenumber zone between SH and H instabilities entirely disappears as soon as odd viscosity increases. 
This ensures the occurrence of a transition from SH to H instability rapidly and in a continuous fashion in the finite wavenumber regime when the odd viscosity increases. 
% This ensures that the transition from subharmonic to harmonic instability takes place rapidly in a continuous fashion in the finite wavenumber regime as long as the odd viscosity increases. 
A comparatively higher inertia force shrinks the unstable region of the SH and H instabilities and consequences more unstable regions of the SH and H waves in an alternative fashion in the marginal curve. Nevertheless, the critical amplitude needed for the first SH instability is always lower compared to the first H instability.

Additionally, shear instability comes into existence together with the gravitational, SH, and H instabilities for a strong interia force with a small angle of inclination. The gravitational and resonance instabilities can be weakened by choosing a higher odd viscosity value, whereas an opposite trend is observed for the shear instability. In view of the above, examining the behavior of surface wave instability of unsteady laminar flow overlying a vibrating inclined plane advances our level of understanding when the time-reversal symmetry breaks in the flow model. A detailed elucidation of the interplay between the time-dependent base state and the odd viscosity emphasizes the importance of considering odd viscosity in the study of flow stability. 

\newpage
\section*{Appendix}
\appendix
\section{Expression of the block tri-diagonal matrix $\mathcal{F}$ and block diagonal matrix $\mathcal{G}$}\label{mat}
 \setcounter{figure}{0}
\begin{align*}
& \mathcal{F}=\begin{bmatrix}
\vdots & \vdots & \vdots&\vdots&\vdots&\vdots&\vdots\\
\dots & (\mathcal{A}+2\mathsf{i}\mathcal{B}) & \mathcal{N}&0&0&0&\dots\\
\dots&\mathcal{N}^{*}&(\mathcal{A}+\mathsf{i}\mathcal{B})&\mathcal{N}&0&0&\dots\\
\dots&0&\mathcal{N}^{*}&\mathcal{A}&\mathcal{N}&0&\dots\\
\dots&0&0&\mathcal{N}^{*}&(\mathcal{A}-\mathsf{i}\mathcal{B})&\mathcal{N}&\dots\\
\dots&0&0&0&\mathcal{N}^{*}&(\mathcal{A}-2\mathsf{i}\mathcal{B})&\dots\\
\vdots&\vdots&\vdots&\vdots&\vdots&\vdots&\vdots
\end{bmatrix}    
\end{align*}
and 
\begin{align*}
& \mathcal{G}=\begin{bmatrix}
\vdots & \vdots & \vdots&\vdots&\vdots&\vdots&\vdots\\
\dots & \mathcal{B} & 0&0&0&0&\dots\\
\dots&0& \mathcal{B}&0&0&0&\dots\\
\dots&0&0&\mathcal{B}&0&0&\dots\\
\dots&0&0&0&\mathcal{B}&0&\dots\\
\dots&0&0&0&0&\mathcal{B}&\dots\\
\vdots&\vdots&\vdots&\vdots&\vdots&\vdots&\vdots
\end{bmatrix} .  
\end{align*}
\section*{Acknowledgment}
SG gratefully acknowledges the financial support from SERB, Department of Science and Technology, Government of India through Award No. MTR/2021/000442.
\section*{Conflict of Interest}
I have no conflict of interest. 

\section*{Declaration of competing interest}
The author declares that he has no known competing financial interests or personal relationships that could have appeared to influence the work reported in this paper.

\section*{Credit authorship contribution statement}
\noindent Md. Mouzakkir Hossain: Conceptualization, Methodology, Software, Writing - original draft, Validation, Formal analysis, Investigation.
\\
Mrityunjoy Saha: Methodology, Formal analysis, Investigation.
\\
Harekrushna Behera: Software, Review \& Editing.
\\
Sukhendu Ghosh: Conceptualization, Software, Review \& Editing.

\bibliographystyle{unsrtnat}
\bibliography{REF}
\end{document}